\documentclass[12pt,a4paper]{article}
\usepackage[dvipdfm]{graphicx,color}
\usepackage{bm}
\usepackage{amsmath,amssymb}
\usepackage{booktabs}
\makeatletter
\def\mbf#1{\mbox{\boldmath ${#1}$}}
\def\Slash{\mathpalette\@Slash}
\def\@Slash#1#2{{\ooalign{\hfil$#1/$\hfil\crcr$#1{#2}$}}}
\makeatother
\begin{document}

\title{Sizable D-term contribution as a signature \\
of $E_6 \times SU(2)_F \times U(1)_A$ SUSY GUT model}
\author{
\centerline{
Nobuhiro~Maekawa$^{1,2}$\footnote{E-mail address: maekawa@eken.phys.nagoya-u.ac.jp}
,~ 
Yu~Muramatsu$^{1}$\footnote{E-mail address: mura@eken.phys.nagoya-u.ac.jp}
~and 
Yoshihiro~Shigekami$^{1}$\footnote{E-mail address: sigekami@eken.phys.nagoya-u.ac.jp}}
\\*[25pt]
\centerline{
\begin{minipage}{\linewidth}
\begin{center}
$^1${\it \normalsize Department of Physics, Nagoya University, Nagoya 464-8602, Japan }  \\*[10pt]
$^2${\it \normalsize Kobayashi Maskawa Institute, Nagoya University, Nagoya 464-8602, Japan }  \\*[10pt]\end{center}
\end{minipage}}
\\*[50pt]}
\date{}
\maketitle
\begin{abstract}
We show that the sizable $D$-term contributions to sfermion mass spectrum can be 
signatures of certain GUT, $E_6\times SU(2)_F\times U(1)_A$ GUT. Note that these $D$-term
contributions deviate the degenerate sfermion masses among different generations in this model. 
This is different from  
the previous works which have argued the $D$-term contributions, which deviate masses 
only between sfermions with different quantum charges, as a signature of GUT with
larger rank unification group. Such $D$-terms are strongly constrained by the FCNC processes
if the SUSY breaking scale is the weak scale. However, in $E_6\times SU(2)_F\times U(1)_A$,
natural SUSY type sfermion mass spectrum is obtained, and if the masses of ${\bf 10}_3$
 sfermions  are larger than $O(1{\rm TeV})$ to realize 126 GeV Higgs and the other sfermion masses are
 $O(10{\rm TeV})$, then sizable $D$-term contribution is allowed. If the deviations by these
 $D$-terms can be observed in future experiments like 100 TeV proton collider or muon collider,
 we may confirm the $E_6\times SU(2)_F\times U(1)_A$ GUT.

\end{abstract}

\section{Introduction}
Grand unified theory (GUT)\cite{GUT} is one of the most promising extensions of the 
standard model (SM).
It unifies not only three gauge interactions in the SM into a single gauge interaction, but also
quarks and leptons into a few multiplets, $\bf 10$ and $\bf\bar 5$ of $SU(5)$, for example.
Moreover, there are experimental supports for both unifications.
For the unification of forces, measured values of three gauge couplings are consistent with 
the unification of gauge interactions quantitatively in the minimal supersymmetric 
(SUSY) SM (MSSM). For the unification of matters in $SU(5)$ GUT, if we assume 
that the $\mbf{10}$ matter fields induce stronger hierarchies of Yukawa couplings than the 
$\bar{\mbf{5}}$ matter fields, measured various hierarchies of quark and lepton masses and 
mixings can be explained qualitatively at the same time\cite{anarchy}. 

In $E_6$ unification\cite{E6,E6neutrino}, the above
assumption for the origin of the hierarchies can be derived\cite{E6matter}. 
As a result, we can obtain
various realistic hierarchies of Yukawa couplings from one basic Yukawa hierarchy which
realizes the hierarchy of up-type quarks. Moreover, if the family symmetry\cite{family}, 
$SU(3)_F$ or
$SU(2)_F$, is introduced, we can obtain a model in which all three generations of quarks 
and leptons can be unified into a single multiplet or two multiplets, and after breaking
the family symmetry and $E_6$ unified symmetry, realistic quark and lepton masses and 
mixings can be realized\cite{E6family}. 
Such models predict a peculiar sfermion mass spectrum in which all sfermions except the third
generation of the $\mbf{10}$ matter $\mbf{10}_3$ have universal sfermion masses. 
This is called modified universal sfermion masses (MUSM). 
 When the mass of $\mbf{10}_3$ is smaller than the other universal sfermion masses, 
 the mass spectrum is nothing but the natural SUSY type sfermion mass 
 spectrum\cite{naturalSUSY}, in which
the SUSY flavor changing neutral currents (FCNC)  processes are suppressed due to large sfermion
masses while the weak scale is stabilized. 

The most difficult problem in SUSY GUT scenario is the doublet-triplet splitting problem\cite{DT}.
One pair of Higgs doublets in the MSSM can be included in $\mbf 5_H$ and ${\bf\bar 5}_H$
with one pair of triplet (colored) Higgs. The mass of triplet Higgs must be larger than
the GUT scale to stabilize the nucleon, while the mass of doublet Higgs must be around the
weak scale. It is difficult to realize such a splitting without finetuning. Several ideas to 
solve this problem have been discussed in various models in the literature. Unfortunately, 
in most of the models, very small parameters are required or the terms which are allowed by 
the symmetry are dropped just by hand. Such a feature is in a sense a finetuning.

If the anomalous $U(1)_A$ gauge symmetry\cite{anomalous} is introduced, 
the doublet-triplet splitting problem
can be solved in a natural assumption that all the interactions which are allowed by the 
symmetry are introduced with $O(1)$ coefficients. Note that all higher dimensional interactions
which are allowed by the symmetry are introduced. Because of this natural assumption, we call
the GUT scenario with the anomalous $U(1)_A$ gauge symmetry ``natural GUT"\cite{naturalGUT,mu}.
 Note that in natural
GUT 
the vacuum expectation value (VEV) of an operator $O_i$ can be determined by its $U(1)_A$ charge 
$o_i$ as
\begin{equation}
\langle O_i\rangle=\left\{\begin{array}{c}  0   \qquad (o_i > 0) \\
                                           \lambda^{-o_i} \qquad (o_i \leq 0)
                         \end{array}\right.,
\end{equation} 
where $\lambda$ is determined from the Fayet-Iliopoulos parameter $\xi$ as 
$\lambda\equiv \xi/\Lambda$.
In this paper, we take $\lambda\sim 0.22$ and adopt the unit in which the cutoff $\Lambda=1$.
This feature is important in solving the 
doublet-triplet splitting problem.

If we consider the $E_6$ GUT with family symmetry and the anomalous $U(1)_A$ gauge symmetry
at the same time, more attractive GUT model with $E_6\times SU(2)_F\times U(1)_A$ gauge symmetry
can be obtained. 
Since the $\mu$ problem is also solved in the natural GUT\cite{mu,non-GUT mu}, we can discuss the SUSY CP problem.
Actually, by imposing the CP symmetry and considering the spontaneous CP violation,
we can solve not only the usual SUSY CP problem but also the new SUSY CP problem on 
chromo-electric dipole moment (CEDM)\cite{CEDM} which is more serious in natural SUSY type sfermion mass
spectrum\cite{SpCP,SpCPneutrino}. 

How to test this interesting SUSY GUT scenario? Since the unification scale is so large that
it is difficult to produce GUT particles directly, it is important to examine various indirect
searches. The most promising candidate for the indirect search is to find the nucleon decay.
In the natural GUT, the nucleon decay via dimension 6 operators is enhanced while the nucleon 
decay via dimension 5 operators is suppressed\cite{naturalGUT}. 
We have proposed how to identify the unification group in the natural GUT by observing the
decay modes of nucleon in Ref. \cite{Proton, ProtonSU2}.

Alternatively, we can test the GUT scenario by measuring sfermion mass spectrum if some
signatures of the GUT appear in the sfermion masses.
For example, if the rank of the unification group is larger than 4, the non-vanishing
$D$-term contributions which are usually flavor blind can be a signature of the GUT scenario
\cite{kawamura}.
The MUSM can be a signature of $E_6\times SU(2)_F$ GUT, since the most serious CEDM 
constraints for the natural SUSY type mass spectrum can be avoided in the scenario by 
spontaneous CP violation. One more interesting test for the $E_6\times SU(2)_F$ GUT scenario
is to observe the non-vanishing $D$-term contributions of the $E_6$ and $SU(2)_F$ gauge symmetry
 to sfermion masses. It is interesting because they spoil the universality of the sfermion 
 masses. 
 Before the LHC found the 126 GeV Higgs\cite{Higgs}, these $D$-term contributions are
 strongly constrained to be small from the various  FCNC processes\cite{yoshioka}.
 However, the stop mass must be larger than 1 TeV in order to realize 126 GeV Higgs, and 
 therefore the other sfermion masses can be O(10 TeV). Since FCNC constraints become much 
 milder when the SUSY breaking scale is O(10 TeV), sizable $D$-term contribution may be
 allowed. 
 
 In this paper, we clarify the $D$-term contributions of $E_6\times SU(2)_F\times U(1)_A$
 GUT model and discuss the FCNC constraint mainly from $\epsilon$ of $K^0\bar K^0$ mixing.
 We will conclude that sizable $D$-term contribution is possible. If the $D$-term
 contributions are sufficiently large and observed by future experiments, for example,
 by the SuperLHC, then we can obtain the precious evidence of the GUT scenario.

\section{$E_6 \times SU(2)_F \times U(1)_A$ SUSY GUT model}
In this section we give a short review on $E_6 \times SU(2)_F \times U(1)_A$ 
SUSY GUT model. Please refer to \cite{E6,E6neutrino} for more detail explanation on the 
model and
the notation for the GUT model in this paper is almost the same as these for the model 
in Ref. \cite{ProtonSU2}.

\subsection{Yukawa matrices for quarks and charged leptons}
Contents of matters and Higgs and their charge assignment are shown in Table 1,
though this model is just an example.
\begin{table}[tbp]
\begin{center}
\begin{tabular}{c|ccccccccccc}
 & $\Psi_a$ & $\Psi_3$ & $F_a$ & $\bar{F}^a$ & $\Phi$ & $\bar{\Phi}$ & $C$ & $\bar{C}$ & $A$ & $Z_3$ &$\Theta$\\
\hline
$E_6$ & ${\bf 27}$ & ${\bf 27}$ & ${\bf 1}$ & ${\bf 1}$ & ${\bf 27}$ & ${\bf \overline{27}}$ &
${\bf 27}$ & ${\bf \overline{27}}$ & ${\bf 78}$ & {\bf 1} & {\bf 1} \\
$SU(2)_F$ & ${\bf 2}$ & ${\bf 1}$ & ${\bf 2}$ & ${\bf \bar{2}}$ & ${\bf 1}$ & ${\bf 1}$ & 
${\bf 1}$ & ${\bf 1}$ & {\bf 1} & {\bf 1} &{\bf 1} \\
$U(1)_A$ & 4 & $\frac{3}{2}$ & -$\frac{3}{2}$ & -$\frac{5}{2}$ & -3 & 1 & -4 & -1 & -$\frac{1}{2}$ & -$\frac{3}{2}$ & -1\\
$Z_6$ & 3 & 3 & 1 & 0 & 0 & 0 & 5 & 0 & 0 & 0 & 0\\
\hline
\end{tabular}
\caption{Field contents and charge assignment under $E_6$$\times SU(2)_F\times U(1)_A\times Z_6$.}
\label{tb:Field contents}
\end{center}
\end{table}
In the model we introduce three $27$ dimensional (fundamental) fields of 
$E_6$ as matters.
$\bf 27$ is decomposed in the $E_6\supset SO(10)\times U(1)_{V'}$ notation 
(and in the [$SO(10)\supset SU(5)\times U(1)_V$] notation) as 
\begin{equation}
\bm{27}=\bm{16}_1[\bm{10}_{1}+\bar{\bm{5}}_{-3}+\bm{1}_{5}]+
\bm{10}_{-2}[\bm{5}_{-2}+\bm{\bar{5}'}_{2}]+\bm{1}'_4[\bm{1}'_0].
\end{equation}
$\bf 27$ of $E_6$ includes not only spinor $\bf 16$ but also vector $\bf 10$ of $SO(10)$.
These $\bf 10$s of $SO(10)$ play an important role in obtaining realistic quark and lepton 
masses and mixings.
Spinor and vector of $SO(10)$ are decomposed in the 
$SU(3)_C \times SU(2)_L \times U(1)_Y$ notation as
\begin{equation}
{\bf 16}\rightarrow \underbrace{q_{L}({\bf 3,2})_{\frac{1}{6}}
+u^c_R({\bf \bar{3},1})_{-\frac{2}{3}}
+e^c_R({\bf 1,1})_1}_{{\bf 10}}+
\underbrace{d^c_R({\bf \bar{3},1})_{\frac{1}{3}}
+l_L({\bf 1,2})_{-\frac{1}{2}}}_{{\bf \bar{5}}}+
\underbrace{\nu^c_R({\bf 1,1})_0}_{{\bf 1}},
\end{equation}
\begin{equation}
{\bf 10}\rightarrow \underbrace{D^c_R({\bf \bar{3},1})_{\frac{1}{3}}
+L_L({\bf 1,2})_{-\frac{1}{2}}}_{{\bf \bar{5}'}}+
\underbrace{\overline{D^c_R}({\bf 3,1})_{-\frac{1}{3}}
+\overline{L_L}({\bf 1,2})_{\frac{1}{2}}}_{{\bf 5}}.
\end{equation}
Matter fields ${\bf 27}_i$ $(i=1,2,3)$ include six $\bf \bar 5$s of $SU(5)$.
Three of six $\bf \bar 5$s become superheavy by developing
the VEVs $\langle\Phi\rangle$, which breaks $E_6$ into $SO(10)$, and $\langle C\rangle$, which
breaks $SO(10)$ into $SU(5)$, through the superpotential
\begin{eqnarray}
W_Y&=&(a\Psi_3\Psi_3+b\Psi_3\bar F^a\Psi_a+c\bar F^a\Psi_a\bar F^b\Psi_b)\Phi+d(\Psi_a,\Phi,\bar \Phi,A,Z_3,\Theta) \nonumber \\
&&+f'\bar F^a\Psi_a\epsilon^{bc}F_b\Psi_cC+g'\Psi_3\epsilon^{ab}F_a\Psi_bC,
\end{eqnarray}
where $a$, $b$, $c$, $f'$ and $g'$ are $O(1)$ coefficients. $d(\Psi_a,\Phi,\bar \Phi,A,Z_3,\Theta)$ 
is a gauge invariant function  of $\Psi_a$, $\Phi$, $\bar\Phi$, $A$, $Z_3$ and $\Theta$, and it 
contributes to $\Psi_1\Psi_2\Phi$.
The other three $\bf \bar 5$s become the SM ${\bf \bar 5}^0_i$ whose
main components 
become
 $({\bf \bar 5}^0_1, {\bf\bar 5}^0_2, {\bf\bar 5}^0_3)\sim ({\bf\bar 5}_1, {\bf\bar 5}'_1, 
 {\bf\bar 5}_2)$.
This is a critical observation in calculating the $D$-term contribution to the sfermion masses.

After developing VEVs $\langle \bar\Phi \Phi\rangle\sim \lambda^2$, 
$\langle \bar CC\rangle\sim \lambda^5$, $\langle A\rangle\sim \lambda^{1/2}$,
$\langle \bar F\rangle\sim (0, \lambda^2)$ and $\langle F\rangle\sim (0,e^{i\rho}\lambda^2)$, 
we can obtain the up-type Yukawa matrix $Y_u$, down-type Yukawa matrix $Y_d$, and charged
lepton Yukawa matrix $Y_e$ at the GUT scale as
\begin{eqnarray}
&Y_u&=\left(
\begin{array}{ccc}
0 & \frac{1}{3}d_q \lambda^5 & 0 \\
-\frac{1}{3}d_q \lambda^5 & c\lambda^4 & b\lambda^2  \\
0 & b\lambda^2 & a 
\end{array}
\right), 
\label{eq:Yu} \\
&Y_d &= \nonumber\\
&&\hspace{-1cm}\left(
\begin{array}{ccc}
-\left(\frac{(bg-af)^2}{ac-b^2}+g^2\right)
\frac{\beta_H}{a} e^{i(2\rho-\delta)}\lambda^6    &
-\frac{bg-af}{ac-b^2}\frac{2}{3}d_5 \beta_H e^{i(\rho-\delta)} \lambda^{5.5} & 
\frac{1}{3}d_q \lambda^5 \\ 
\left( -\frac{d_q}{3}-\frac{bg-af}{ac-b^2} \frac{b \frac{2}{3}d_5}{g} \right) \lambda^5  &
\left(f \beta_H e^{i(\rho-\delta)} -\frac{(\frac{2}{3} d_5)^2}{ac-b^2}\frac{ab}{g}e^{-i\rho}  
 \right) \lambda^{4.5} & 
\frac{cg-bf}{g} \lambda^4 \\
-\frac{bg-af}{ac-b^2} \frac{a \frac{2}{3}d_5}{g} \lambda^3 &
\left(g \beta_H e^{i(\rho-\delta)} -\frac{(\frac{2}{3} d_5)^2}{ac-b^2}\frac{a^2}{g}e^{-i\rho}  
 \right) \lambda^{2.5} & 
\frac{bg-af}{g} \lambda^2
\end{array} \right), \nonumber\\
\label{eq:Yd}\\
&Y_e& =\left(
\begin{array}{ccc}
-\left(\frac{(bg-af)^2}{ac-b^2}+g^2\right) 
\frac{\beta_H}{a} e^{i(2\rho-\delta)} \lambda^6 & 
d_l \lambda^5 & 0 \\
0 & f \beta_H e^{i(\rho-\delta)} \lambda^{4.5} & g \beta_H e^{i(\rho-\delta)} \lambda^{2.5}   \\
-d_l \lambda^5 & \frac{cg-bf}{g} \lambda^4 &\frac{bg-af}{g} \lambda^2
\end{array}\right),
\label{eq:Ye}
\end{eqnarray}
where $a,b,c,d_q,d_5,d_l,f,g$ and $\beta_H$ are real $O(1)$ coefficients, 
$\rho$ and $\delta$ are
$O(1)$ phases, and $\lambda\sim 0.22$ is taken to be Cabibbo angle.
In this paper, we begin our arguments from these Yukawa matrices which have only 9 real
 parameters and 2 CP phases.

\subsection{Mass spectrum of sfermions}
The sfermion masses can be obtained from the SUSY breaking potential
\begin{eqnarray}
V_{SB}&=&m_0^2|\Psi_a|^2+m_3^2|\Psi_3|^2+m_{11}^2|\epsilon^{ab}\Psi_a F_b|^2
  +m_{22}^2|\Psi_a\bar F^a|^2 \nonumber\\
&&\hspace{-1cm} +(m_{23}^2\Psi_3^\dagger\Psi_a\bar F^a+m_{13}^2\lambda^5\Psi_3^\dagger\epsilon^{ab}\Psi_a \bar F_b^\dagger
  +m_{12}^2\lambda^5(\Psi_a\bar F^a)^\dagger\epsilon^{bc}\Psi_b \bar F_c^\dagger+h.c.)\\
  &&+m^2(\Phi^\dagger\Psi^{a\dagger}\Psi_a\Phi)
  +(m_{12}^{\prime 2}\lambda^2\bar C|\Psi_a|^2\bar\Phi^\dagger
  +m_{23}^{\prime 2}\lambda^2\bar C\Psi_3^\dagger\Psi_a\bar F^a\bar\Phi^\dagger+h.c.), \nonumber
\end{eqnarray}
where the terms in the last line give the mass terms which do not respect $E_6$ symmetry because
the VEVs $\langle\Phi\rangle$ and $\langle C\rangle$ break $E_6$ symmetry.
The $D$-term contributions are written as
\begin{equation}
\Delta\tilde{m}_{\psi}^2
=\sum_I Q_I(\psi)D_I,
\end{equation}
where $D_I$ is the squared gauge coupling times the $D$-term of 
$U(1)_{V'} (I=6)$, $U(1)_V (I=10)$, $U(1)_F (I=F)$ and 
$U(1)_A (I=A)$, and 
$Q_I(\psi)$ is the $U(1)$ charge of the field $\psi$.
Here $U(1)_F$ is the Cartan part of $SU(2)_F$. 
As a result, sparticle masses  for $\bf \bar 5$, 
$\bf \bar 5'$ and $\bf 10$ of $SU(5)$ are 
\begin{eqnarray}
\tilde{m}^2_{\bar{5}} &= \left(
\begin{array}{ccc}
m_0^2+\lambda^4m_{11}^2 &\lambda^9m_{12}^2 & \lambda^7m_{13}^2  \\
\lambda^9m_{12}^2 & m_0^2+\lambda^4m_{22}^2 & \lambda^2m_{23}^2 \\
\lambda^7m_{13}^2 &\lambda^2m_{23}^2 & m_3^2
\end{array}
\right)+
D_6\left(
\begin{array}{ccc}
1 & &  \\
 & 1 &  \\
 & & 1 
\end{array}
\right)\nonumber \\
&\hspace{-1cm}+D_{10}\left(
\begin{array}{ccc}
-3 & &  \\
 & -3 &  \\
 & & -3 
\end{array}
\right) 
+D_F\left(
\begin{array}{ccc}
1 & &  \\
 & -1 &  \\
 & & 0 
\end{array}
\right)+
D_A\left(
\begin{array}{ccc}
4 & &  \\
 & 4 &  \\
 & & \frac{3}{2} 
\end{array}
\right), \\
\tilde{m}^2_{\bar{5}'} &= \left(
\begin{array}{ccc}
m_0^2+\lambda^2m^2+\lambda^4m_{11}^2 &\lambda^9m_{12}^2 & \lambda^7m_{13}^2  \\
\lambda^9m_{12}^2 & m_0^2+\lambda^2m^2+\lambda^4m_{22}^2 & \lambda^2m_{23}^2 \\
\lambda^7m_{13}^2 &\lambda^2m_{23}^2 & m_3^2
\end{array}
\right)
 \nonumber \\
&+D_6\left(
\begin{array}{ccc}
-2 & &  \\
 & -2 &  \\
 & & -2 
\end{array}
\right)+D_{10}\left(
\begin{array}{ccc}
2 & &  \\
 & 2 &  \\
 & & 2 
\end{array}
\right)
+D_F\left(
\begin{array}{ccc}
1 & &  \\
 & -1 &  \\
 & & 0 
\end{array}
\right) \\
&+D_A\left(
\begin{array}{ccc}
4 & &  \\
 & 4 &  \\
 & & \frac{3}{2} 
\end{array}
\right),&
\nonumber\\
\tilde{m}^2_{10} &= \left(
\begin{array}{ccc}
m_0^2+\lambda^4m_{11}^2 &\lambda^9m_{12}^2 & \lambda^7m_{13}^2  \\
\lambda^9m_{12}^2 & m_0^2+\lambda^4m_{22}^2 & \lambda^2m_{23}^2 \\
\lambda^7m_{13}^2 &\lambda^2m_{23}^2 & m_3^2
\end{array}
\right)+
D_6\left(
\begin{array}{ccc}
1 & &  \\
 & 1 &  \\
 & & 1 
\end{array}
\right) \nonumber \\
&+D_{10}\left(
\begin{array}{ccc}
1 & &  \\
 & 1 &  \\
 & & 1 
\end{array}
\right)
+D_F\left(
\begin{array}{ccc}
1 & &  \\
 & -1 &  \\
 & & 0 
\end{array}
\right)+
D_A\left(
\begin{array}{ccc}
4 & &  \\
 & 4 &  \\
 & & \frac{3}{2} 
\end{array}
\right), \label{10}
\end{eqnarray}
where the contribution of the term $m^2\Phi^\dagger\Psi^{a\dagger}\Psi_a\Phi$ to 
$|{\bf 16}_{\Psi_a}|^2$ is included in $m_0^2$ by redefinition of $m_0^2$.
Then, the sfermion mass matrix for SM $\bf \bar 5$ fields, which are represented as
$({\bf\bar 5}^0_1,{\bf\bar 5}^0_2,{\bf\bar 5}^0_3)\sim
({\bf\bar 5}_1,{\bf\bar 5'}_1,{\bf\bar 5}_2)$, 
becomes
\begin{align}
\tilde{m}^2_{\bar{5}^0} 
&\sim \left(
\begin{array}{ccc}
m_0^2+\lambda^4m_{11}^2 &\lambda^{5.5}m_{12}^{\prime 2} & \lambda^9m_{12}^2  \\
\lambda^{5.5}m_{12}^{\prime 2} & m_0^2+\lambda^2m^2+\lambda^4m_{11}^2 & 
\lambda^{7.5}m_{23}^{\prime 2} \\
\lambda^9m_{12}^2 &\lambda^{7.5}m_{23}^{\prime 2} & m_0^2+\lambda^4m_{22}^2
\end{array}
\right) \nonumber \\
&+D_6\left(
\begin{array}{ccc}
1 & &  \\
 & -2 &  \\
 & & 1 
\end{array}
\right)+
D_{10}\left(
\begin{array}{ccc}
-3 & &  \\
 & 2 &  \\
 & & -3 
\end{array}
\right) \label{bar5}\\
&+D_F\left(
\begin{array}{ccc}
1 & &  \\
 & 1 &  \\
 & & -1 
\end{array}
\right)+
D_A\left(
\begin{array}{ccc}
4 & &  \\
 & 4 &  \\
 & & 4
\end{array}
\right). \nonumber
\end{align}
Moreover, the contributions from the sub-leading components of ${\bf\bar 5}^0_i$ become
\begin{equation}
\Delta\tilde{m}^2_{\bar{5}^0} 
\sim (m_0^2-m_3^2)\left(
\begin{array}{ccc}
\lambda^6 &\lambda^{5.5} & \lambda^5 \\
\lambda^{5.5} & \lambda^5 & \lambda^{4.5} \\ 
\lambda^5 &\lambda^{4.5} &\lambda^4
\end{array}
\right).
\end{equation}
These sfermion mass matrices give interesting predictions of $E_6\times SU(2)_F\times U(1)_A$
GUT, though the terms which are suppressed by power of $\lambda$ are strongly dependent
on the explicit model. In the next section, we discuss how to obtain the GUT information
from sfermion mass spectrum.

\section{Signatures of $E_6\times SU(2)_F\times U(1)_A$ GUT from sfermion mass spectrum}
Suppose that all sfermion masses are measured by experiments in future. 
In this section, we discuss the signatures of the GUT in sfermion mass spectrum at the GUT
scale. We have to use renormalization groups (RGs) to obtain the sfermion mass spectrum at the
weak scale, but we do not discuss the effects in this section. 
The constraints from the FCNC processes will be discussed in the next section.

If the observed sfermion mass spectrum is the MUSM as
\begin{equation}
\tilde m_{10}^2\sim\left(\begin{array}{ccc}
                          m_0^2 & & \\
                          & m_0^2 & \\
                          && m_3^2
                         \end{array}\right),\quad
\tilde m_{\bar 5^0}^2\sim\left(\begin{array}{ccc}
                          m_0^2 & & \\
                          & m_0^2 & \\
                          && m_0^2
                         \end{array}\right),
\end{equation}
$E_6\times SU(2)_F$ GUT is strongly implied.
Of course, the MUSM is nothing but a natural SUSY type sfermion mass spectrum, which
are predicted by a lot of models. 
However, generically natural SUSY type sfermion mass spectrum is suffering from
the CEDM problem\cite{CEDM}, and there are few models in which the problem can be solved in a natural
way. We would like to emphasize that 
the CEDM problem can be solved in the $E_6\times SU(2)_F\times U(1)_A$ GUT by spontaneous 
CP violation in a non-trivial way\cite{SpCP}.

In order to obtain more specific signatures of the $E_6\times SU(2)_F\times U(1)_A$ GUT,
we study the $D$-term contributions. 
For a while, we neglect the terms which are suppressed by power of $\lambda$.
We will discuss these terms later.
Then, the mass matrices of $\tilde m_{10}^2$ and $\tilde m_{\bar 5^0}^2$ are rewritten as
\begin{eqnarray}
\tilde m_{10}^2&=&(m_0^2+D_6+D_{10}+D_F+4D_A){\bm 1_{3 \times 3}} \nonumber \\
&+&\left(
\begin{array}{ccc}
0 & &  \\
 & -2D_F &  \\
 & & -D_F-\frac{5}{2}D_A+m_3^2-m_0^2
\end{array}
\right) \nonumber \\
&\equiv& m_{10,0}^2{\bm 1_{3 \times 3}} 
+\left(
\begin{array}{ccc}
0 & &  \\
 & \Delta m^2_{10,2} &  \\
 & & \Delta m^2_{10,3}
\end{array}
\right), \\
\tilde m_{\bar 5^0}^2
&=&(m_0^2+D_6-3D_{10}+D_F+4D_A){\bm 1_{3 \times 3}} \nonumber \\
&+&\left(
\begin{array}{ccc}
0 & &  \\
 & -3D_6+5D_{10} &  \\
 & & -2D_F
\end{array}
\right) \nonumber \\
&\equiv& m_{\bar 5^0,0}^2{\bm 1_{3 \times 3}} 
+\left(
\begin{array}{ccc}
0 & &  \\
 & \Delta m^2_{\bar 5,2} &  \\
 & & \Delta m^2_{\bar 5,3}
\end{array}
\right), 
\end{eqnarray}
where ${\bm 1_{3 \times 3}}$ is a $3\times 3$ unit matrix.
A non-trivial prediction of this model is $\Delta m^2_{10,2}=\Delta m^2_{\bar 5,3}$.
If this relation is observed, we obtain a strong evidence for this model and can know
the $D_F$. 
The $D_6$ and  $D_{10}$ can be determined if $m_{10,0}^2-m_{\bar 5^0,0}^2$ and 
$\Delta m_{\bar 5,2}^2$ are observed.
If these small modifications from the MUSM and $\Delta m^2_{10,2}=\Delta m^2_{\bar 5,3}$
are observed, we think that the $E_6\times SU(2)_F\times U(1)_A$ model can be established.

How large $D$-terms are allowed? If these $D$-terms are very small, it may become difficult to
measure them, and if these $D$-terms are large, the FCNC constraints cannot be satisfied.
In the next section, we study the constraints to the $D$-terms from the FCNC processes, 
especially from $\epsilon$ parameter in $K^0\bar K^0$ mixing, which gives the strongest 
constraints.

\section{FCNC constraints to $D$-terms}
In this section, we focus on the natural SUSY type sfermion masses, i.e., $m_0\gg m_3$,
because the FCNC constraints become milder and the sizable $D$-term may be allowed. 
Therefore, we fix $\Delta m_{10,3}^2=m_0^2$. 
To obtain 126 GeV Higgs, $m_3$ must be larger than 1 TeV. Since the smaller 
$m_3$ leads to be more natural, we take $m_3\sim O(1{\rm TeV})$.
In the literature, the upper bound for the ratio $m_0/m_3$ has been studied, which is
derived from the requirement of the positivity of stop mass square to be 
roughly 5 through two loop RG contribution\cite{naturalSUSY}.
Therefore, we expect that
$m_0$ is $O(10 {\rm TeV})$.
And we consider the constraints from
the CP violating parameter $\epsilon$ in $K^0\bar K^0$ mixing because they are 
the most strong constraints in this situation. 
In this paper, we do not argue the upper bound of $m_0/m_3$ explicitly, 
because larger stop mass
can always satisfy the positivity and the upper bound is dependent on the explicit models
between the GUT scale and the SUSY breaking scale.

Since we calculate constraints from the FCNC processes with the mass eigenstates of quarks and 
leptons, 
we need diagonalizing matrices which make Yukawa matrices diagonal as
\begin{align}
\psi_{L i} (Y_{\psi})_{ij} \psi_{Rj}^c &= 
(L_{\psi}^{\dagger} \psi_{L})_i (L_{\psi}^T Y_{\psi} R_{\psi})_{ij} 
(R_{\psi}^{\dagger}\psi_{R}^c)_j \\
 &\equiv\psi'_{L i} (Y_{\psi}^D)_{ij} \psi'^c_{Rj}, \nonumber
\end{align}
where $\psi$ is a flavor eigenstate, $\psi'$ is a mass eigenstate and $Y_{\psi}^D$ is a
diagonalized matrix of $\psi$.
We summarize the detail expression of these diagonalizing matrices with the explicit $O(1)$
coefficients in Appendix A.
Here we roughly show the diagonalizing matrices for up-type quark, down-type quark and charged lepton without $O(1)$ coefficients  as 
\begin{equation}
L_u \sim \left(
\begin{array}{ccc}
1 & \frac{1}{3} \lambda & 0 \\
\frac{1}{3} \lambda & 1 & \lambda^2  \\
\frac{1}{3} \lambda^3 & \lambda^2 & 1 
\end{array}
\right),R_u \sim \left(
\begin{array}{ccc}
1 & \frac{1}{3} \lambda & 0 \\
\frac{1}{3} \lambda & 1 & \lambda^2  \\
\frac{1}{3} \lambda^3 & \lambda^2 & 1 
\end{array}
\right),
\label{eq:diagonalizing up}
\end{equation}

\begin{equation}
L_d \sim \left(
\begin{array}{ccc}
1 & (\frac{2}{3}+i\frac{4}{27}) \lambda & \frac{1}{3} \lambda^3 \\
(\frac{2}{3}+i\frac{4}{27}) \lambda & 1 & \lambda^2  \\
(\frac{2}{3}+i\frac{4}{27}) \lambda^3 & \lambda^2 & 1 
\end{array}
\right),R_d \sim \left(
\begin{array}{ccc}
1 & \frac{2}{3}(1+i) \lambda^{0.5} & \frac{2}{3} \lambda \\
\frac{2}{3}(1+i) \lambda^{0.5} & 1 & (1+i)\lambda^{0.5}  \\
\frac{2}{3}(1+i) \lambda & (1+i)\lambda^{0.5} & 1 
\end{array}
\right),
\end{equation}

\begin{equation}
L_e \sim \left(
\begin{array}{ccc}
1 & (1+i)\lambda^{0.5} & 0 \\
(1+i)\lambda^{0.5} & 1 & (1+i)\lambda^{0.5}  \\
\lambda & (1+i)\lambda^{0.5} & 1 
\end{array}
\right),R_e \sim \left(
\begin{array}{ccc}
1 & \lambda & \lambda^3 \\
\lambda & 1 & \lambda^2  \\
\lambda^3 & \lambda^2 & 1 
\end{array}
\right),
\label{eq:diagonalizing electoron}
\end{equation}

\begin{equation}
L_{\nu} \sim \left(
\begin{array}{ccc}
1 & (1+i)\lambda^{0.5} & (1+i)\lambda \\
(1+i)\lambda^{0.5} & 1 & (1+i)\lambda^{0.5}  \\
(1+i)\lambda & (1+i)\lambda^{0.5} & 1 
\end{array}
\right).
\label{eq:diagonalizing neutrino}
\end{equation}
We have two types for the diagonalizing matrix $U$ as 
\begin{equation}
U = \begin{cases}
    U_{\text{CKM-type}}\equiv \left(
\begin{array}{ccc}
1 & a_{12}\lambda & a_{13}\lambda^3  \\
a_{21}\lambda & 1 & a_{23}\lambda^2  \\
a_{31}\lambda^3 & a_{32}\lambda^2 & 1
\end{array}
\right)
& (\text{for $\bm{10}$ of $SU(5)$ multiplet}) \\
    U_{\text{MNS-type}}\equiv \left(
\begin{array}{ccc}
b_{11} & b_{12}\lambda^{0.5} & b_{13}\lambda  \\
b_{21}\lambda^{0.5} & b_{22} & b_{23}\lambda^{0.5}  \\
b_{31}\lambda & b_{32}\lambda^{0.5} & b_{33}
\end{array}
\right) & (\text{for ${\bf \bar 5}$ of $SU(5)$ multiplet})
  \end{cases},
\end{equation}
where $a_{ij}$ and $b_{ij}$ are complex $O(1)$ coefficients.
The mass insertion parameters defined as 
\begin{equation}
(\delta^{\psi}_{ij})_{\Gamma \Gamma}\equiv
\frac{(U^{\dagger}_{\psi_\Gamma}\tilde{m}^2_{\psi_\Gamma}U_{\psi_\Gamma})_{ij}}
{m^2_{\tilde{\psi}}}, \quad (\Gamma=L,R)
\end{equation}
can be calculated as
\begin{align}
&(\delta^{\psi}_{ij})_{\Gamma \Gamma}= \\
&\left(\begin{array}{ccc}
\cdots & a_{21}^*\lambda \Delta m^2_{10,2} +a_{31}^* a_{32} \lambda^5 \Delta m^2_{10,3} & 
(a_{21}^* a_{23} \Delta m^2_{10\ 2}+a_{31}^*  \Delta m^2_{10,3})\lambda^3 \\
\cdots & \cdots & 
(a_{23} \Delta m^2_{10,2}+a_{32}^*  \Delta m^2_{10,3})\lambda^2 \\
\cdots & \cdots & \cdots
\end{array}
\right) / m_{\tilde \psi}^2, \nonumber
\end{align}
\begin{align}
&(\delta^{\psi}_{ij})_{\Gamma \Gamma}= \\
&\left(\begin{array}{ccc}
\cdots & b_{21}^*b_{22}\lambda^{0.5} \Delta m^2_{\bar 5,2} +b_{31}^* b_{32} \lambda^{1.5} \Delta m^2_{\bar 5,3} & 
(b_{21}^* b_{23} \Delta m^2_{\bar 5,2}+b_{31}^* b_{33} \Delta m^2_{\bar 5,3})\lambda \\
\cdots & \cdots & 
(b_{22}^* b_{23} \Delta m^2_{\bar 5,2}+b_{32}^* b_{33} \Delta m^2_{\bar 5,3})\lambda^{0.5} \\
\cdots & \cdots & \cdots
\end{array}
\right) / m_{\tilde \psi}^2, \nonumber
\end{align}
for ${\bf 10}$ fields and $\bf\bar 5$ fields, respectively. 
In Appendix \ref{sec:mass insertions}, we show each mass insertion in this model with
explicit $O(1)$ coefficients.

\begin{figure}[htb]
 \centering
 \includegraphics[width=12cm,clip]{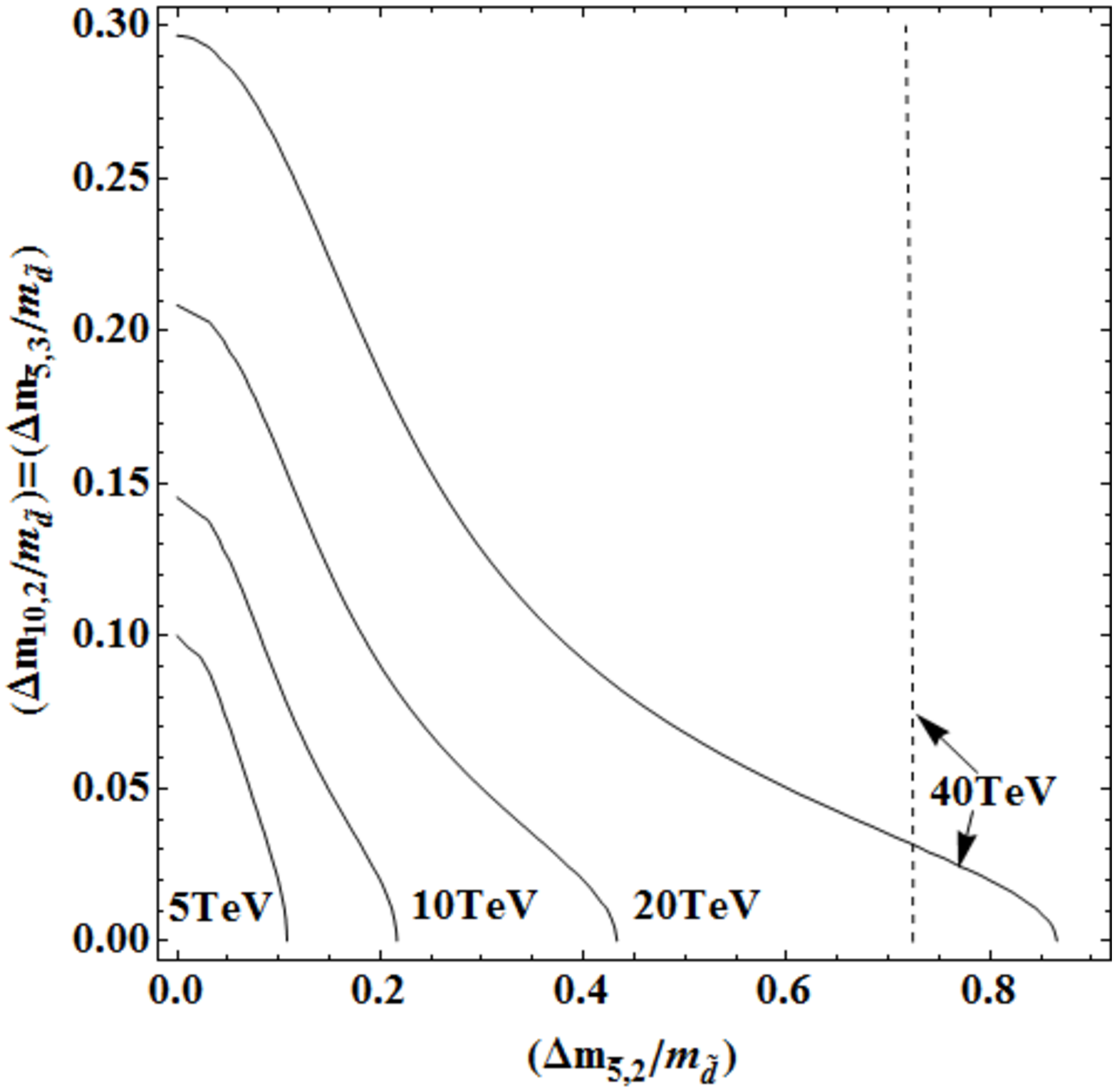}
\caption{Allowed region in 
 $(\Delta m_{\bar 5,2}/m_{\tilde d},\  
 \Delta m_{10,2}/m_{\tilde d}=\Delta m_{\bar 5,3}/m_{\tilde d})$ space. 
The allowed region for the condition of 
$\sqrt{|{\rm Im}(\delta_{12}^d)_{LL}(\delta_{12}^d)_{RR}|}$
is obtained below the solid lines for various $m_{\tilde d}=$ 5 TeV, 10 TeV, 
20 TeV and 40 TeV. The allowed region for $\sqrt{|{\rm Im}(\delta_{12}^d)_{RR}^2|}$
is the left side of the dotted line for $m_{\tilde d}=40$ TeV. The other conditions
are satisfied in the allowed region for 
$\sqrt{|{\rm Im}(\delta_{12}^d)_{LL}(\delta_{12}^d)_{RR}|}$.
}
\label{Fig1}
\end{figure}

Let us calculate the constraints from the $\epsilon$ parameter in $K^0\bar K^0$ mixing.
We use the constraints for $(\delta_{12}^d)_{LL}$ and $(\delta_{12}^d)_{RR}$ as
\begin{eqnarray}
&\sqrt{|{\rm Im}(\delta_{12}^d)_{LL}^2|}&<2.9\times 10^{-3}
\left(\frac{m_{\tilde d}}{500 {\rm GeV}}\right), \\
&\sqrt{|{\rm Im}(\delta_{12}^d)_{RR}^2|}&<2.9\times 10^{-3}
\left(\frac{m_{\tilde d}}{500 {\rm GeV}}\right), \\
&\sqrt{|{\rm Im}(\delta_{12}^d)_{LL}(\delta_{12}^d)_{RR}|}&<1.1\times 10^{-4}
\left(\frac{m_{\tilde d}}{500 {\rm GeV}}\right), 
\end{eqnarray}
which are obtained by requiring that the SUSY contribution $\epsilon_{SUSY}$ is smaller 
 $\epsilon_{SM}$\cite{FCNC}. These parameters can roughly be calculated as
\begin{eqnarray}
(\delta_{12}^d)_{LL}&\sim& \left(\frac{2}{3}+i\frac{4}{27}\right)\left(\lambda\frac{\Delta m_{10,2}^2}{m_{\tilde d}^2}
+\lambda^5\frac{\Delta m_{10,3}^2}{m_{\tilde d}^2}\right) \\
(\delta_{12}^d)_{RR}&\sim& \frac{2}{3}(1+i)\left(\lambda^{0.5}\frac{\Delta m_{\bar 5,2}^2}{m_{\tilde d}^2}+\lambda^{1.5}\frac{\Delta m_{\bar 5,3}^2}{m_{\tilde d}^2}\right).
\end{eqnarray}
By taking $\Delta m_{10,3}^2=m_0^2=m_{\tilde d}^2$, we can obtain the allowed region in 
$(\Delta m_{\bar 5,2}, \Delta m_{\bar 10,2}=\Delta m_{\bar 5,3})$ space, which
 is shown in Fig. \ref{Fig1}. Note that 
 $\Delta m_{10,2}=\Delta m_{\bar 5,3}$ is one of the predictions
 in $E_6\times SU(2)_F\times U(1)_A$ model.
Roughly, if $m_0$ is $O(10 {\rm TeV})$, $\Delta m$ is allowed to be $O(1 {\rm TeV})$ which
is nothing but the scale of $m_3$. 
Interestingly, the constraint to $\Delta m_{\bar 5,2}$ is milder, and sizable 
$\Delta m_{\bar 5,2}$
can be allowed. Since the $E$-twisting structure $({\bf\bar 5}_1,{\bf\bar 5}'_1,{\bf\bar 5}_2)$
is important to obtain the non-vanishing  $\Delta m_{\bar 5,2}$, this can be a critical signature
of $E_6\times SU(2)_F\times U(1)_A$ model. 
On the other hand, the constraint to $\Delta m_{10,2}=\Delta m_{\bar 5,3}$ is stronger, because
this contributes to $(\delta_{12}^d)_{LL}$ and $(\delta_{12}^d)_{RR}$ at the same time.

In the above arguments, we neglected the contributions to the sfermion
masses which are suppressed by power of $\lambda$  in Eqs. (\ref{10}) and (\ref{bar5}).
All these terms except $\lambda^2 m^2$ term in Eq. (\ref{bar5}) can be neglected in the
above arguments. However, $\lambda^2m^2$ term gives non-vanishing
$\Delta m_{\bar 5,2}/m_{\tilde d}$, which becomes $O(\lambda)$ if $m\sim m_0$. 
The FCNC constraints in this situation can be easily extracted from  Fig. \ref{Fig1}.

\section{Discussions and summary}
We have shown that the sizable $D$-term contributions to sfermion mass spectrum can be 
signatures of certain GUT, $E_6\times SU(2)_F\times U(1)_A$ GUT. Note that these $D$-term
contributions deviate the degenerate sfermion masses among different generations in this model. 
This point is one large difference between our work and 
the previous works which have argued the $D$-term contributions\cite{kawamura}, 
which deviate masses 
only between sfermions with different quantum charges, as a signature of GUT with
larger rank unification group. Such $D$-terms are strongly constrained by the FCNC processes
if the SUSY breaking scale is the weak scale. However, in $E_6\times SU(2)_F\times U(1)_A$,
natural SUSY type sfermion mass spectrum is obtained, and if the masses of ${\bf 10}_3$
 sfermions  are larger than $O(1{\rm TeV})$ to realize 126 GeV Higgs and 
 the other sfermion masses are
 $O(10{\rm TeV})$, then sizable $D$-term contribution is allowed. 
A novel relation 
$\tilde m_{{\bf\bar 5}_3}^2-\tilde m_{{\bf\bar 5}_1}^2=\tilde m_{{\bf 10}_2}^2-\tilde m_{{\bf 10}_1}^2$ is predicted.
 If the deviations by these
 $D$-terms can be observed in future experiments like 100 TeV proton collider or muon collider,
 we may confirm the $E_6\times SU(2)_F\times U(1)_A$ GUT.
 
Of course, we have to consider the renormalization group (RG) effects in the estimation 
generically, though we have not considered this effect in this paper. However, if the gravitino
mass is $O(100{\rm TeV})$ to solve the gravitino problem and SUSY breaking parameters in the
MSSM are $O(1{\rm TeV})$ to solve the hierarchy problem, the mirage scale, at which
anomaly mediation contribution can cancel the RG effect, can be the SUSY breaking scale.
As a result, we can observe the GUT signature through the gravity mediation contribution 
directly at the SUSY breaking scale\cite{takayama}.

Since the GUT scale is much larger than the TeV scale which we can reach by experiments,
it is important to consider how to test the GUT scenario. We have discussed 
the $D$-term contributions which are dependent on generations, and 
they can be a promising signatures of $E_6\times SU(2)_F\times U(1)_A$ GUT scenario.

\section{Acknowledgement}
N.M. is supported in part by Grants-in-Aid for Scientific Research from MEXT of 
Japan. This work is partially supported by the Grand-in-Aid for Nagoya University
Leadership Development Program for Space Exploration and Research Program from the MEXT 
of Japan.

\appendix

\section{The coefficients of diagonalizing matrices (in leading order)}
\label{sec:coefficients}
In Appendix A in Ref. \cite{ProtonSU2}, we show how to diagonalize the 
$3 \times 3$ matrix $Y_{ij}$.
Here we show the diagonalizing matrices for up-type quark, down-type quark and 
charged lepton.
The diagonalizing matrices $L_{\psi}$ and $R_{\psi}$ come from mixing angles 
$s_{ij}^{\psi L/R} \equiv \sin\theta_{ij}^{\psi L/R}e^{i\chi_{ij}^{\psi L/R}}$ and 
$c_{ij}^{\psi L/R} \equiv \cos\theta_{ij}^{\psi L/R}$.
In our calculation we use the approximation that the mixing angles are small, i.e.,  
$|s_{ij}^{\psi L/R}| \sim |\theta_{ij}^{\psi L/R}| \ll 1$ 
$(s_{ij}^{\psi L/R} \sim \theta_{ij}^{\psi L/R} e^{i\chi_{ij}^{\psi L/R}})$ and
$c_{ij}^{\psi L/R} \simeq 1$.
In this approximation the diagonalizing matrices are 
\begin{equation}
L_{\psi} \simeq  \begin{pmatrix}
   1 & s_{12}^{\psi L*} & s_{13}^{\psi L*} \\
   -s_{12}^{\psi L} & 1 & s_{23}^{\psi L*} \\
   -s_{13}^{\psi L} + s_{23}^{\psi L}s_{12}^{\psi L} & -s_{23}^{\psi L} & 1
  \end{pmatrix}, 
\end{equation}
\begin{equation}
R_{\psi} \simeq  \begin{pmatrix}
   1 & s_{12}^{\psi R} & s_{13}^{\psi R} \\
   -s_{12}^{\psi R*} & 1 & s_{23}^{\psi R} \\
   -s_{13}^{\psi R*} + s_{23}^{\psi R*}s_{12}^{\psi R*} & -s_{23}^{\psi R*} & 1
  \end{pmatrix}.
\end{equation}

From Eq. (\ref{eq:Yu}), the mixing angles for up-type quark are calculated as 
\begin{equation}
s_{23}^{uL}=s_{23}^{uR*} \simeq \frac{b}{a}\lambda^2 \equiv R_{23}^{uL}\lambda^2,\,
s_{13}^{uL}=s_{13}^{uR*} \simeq 0,\,
s_{12}^{uL}=-s_{12}^{uR*} \simeq \frac{\frac{1}{3}ad_q}{ac-b^2}\lambda \equiv 
\frac{1}{3}R_{12}^{uL}\lambda.
\end{equation}
From Eq. (\ref{eq:Yd}), the mixing angles for down-type quark are calculated as 
\begin{equation}
s_{23}^{dL} \simeq \frac{cg-bf}{bg-af}\lambda^2 \equiv R_{23}^{dL}\lambda^2,
s_{13}^{dL} \simeq \frac{1}{3} \frac{d_q g}{bg-af}\lambda^3 \equiv 
\frac{1}{3}R_{13}^{dL} \lambda^3,
\end{equation}
\begin{eqnarray}
&&s_{12}^{dL} \simeq -\frac{2}{3}\frac{(bg-af)^2d_5}{(ac-b^2)\{f(bg-af)-g(cg-bf)\}}\lambda \nonumber \\
&&\hspace{2cm}+\frac{4}{27} \frac{a^2 d_q d_5^2 }{(ac-b^2)\{f(bg-af)-g(cg-bf)\}\beta_H}e^{-i(2\rho-\delta)}\lambda \nonumber \\
 &&\hspace{0.8cm}\equiv (\frac{2}{3}R_{12}^{dL}+\frac{4}{27}I_{12}^{dL}e^{-i(2\rho-\delta)})\lambda, \nonumber
\end{eqnarray}
\begin{eqnarray}
&&s_{23}^{dR*} \simeq \frac{g^2\beta_H}{bg-af}e^{i(\rho-\delta)}\lambda^{0.5}-
\frac{4}{9}\frac{d_5^2 a^2}{(ac-b^2)(bg-af)}e^{-i\rho}\lambda^{0.5} \\
&&\hspace{0.8cm}\equiv I_{23}^{dR}e^{i(\rho-\delta)}\lambda^{0.5}-\frac{4}{9}I_{23}^{\prime dR}e^{-i\rho}\lambda^{0.5},\, 
s_{13}^{dR*} \simeq -\frac{2}{3}\frac{ad_5}{ac-b^2}\lambda\equiv \frac{2}{3}R^{dR}_{13}\lambda,
\nonumber
\end{eqnarray}
\begin{equation}
s_{12}^{dR*} \simeq \frac{2}{3} \frac{d_5(bg-af)}{\{f(bg-af)-g(cg-bf)\}\beta_H}e^{-i(\rho-\delta)}
\lambda^{0.5} \equiv \frac{2}{3}I_{12}^{dR}e^{-i(\rho-\delta)}\lambda^{0.5}. \nonumber
\end{equation}
From Eq. (\ref{eq:Ye}), the mixing angles for charged lepton are calculated as 
\begin{equation}
s_{23}^{eL} \simeq \frac{g^2 \beta_H}{bg-af}e^{i(\rho-\delta)}\lambda^{0.5} 
\equiv I_{23}^{dR}e^{i(\rho-\delta)}\lambda^{0.5},\,s_{13}^{eL} \simeq 0,
\end{equation}
\begin{equation}
s_{12}^{eL} \simeq \frac{d_l(bg-af)}{\beta_H\{f(bg-af)-g(cg-bf)\}}e^{-i(\rho-\delta)}\lambda^{0.5} 
\equiv I_{12}^{eL}e^{-i(\rho-\delta)}\lambda^{0.5} \nonumber
\end{equation}
\begin{equation}
s_{23}^{eR*}\simeq s_{23}^{dL}\equiv R_{23}^{dL}\lambda^2,\,s_{13}^{eR*}\simeq -\frac{d_l g}{bg-af}
\lambda^3 \equiv R_{13}^{eR}\lambda^3,
\end{equation}
\begin{equation}
s_{12}^{eR*}\simeq \frac{d_l g^2}{\{f(bg-af)-g(cg-bf)\}}\lambda \equiv R_{12}^{eR}\lambda. \nonumber
\end{equation}
The diagonalizing matrices for up-type quark, down-type quark and charged lepton are 
calculated as
\begin{equation}
L_u \sim \left(
\begin{array}{ccc}
1 & \frac{1}{3} R_{12}^{uL}\lambda & 0 \\
-\frac{1}{3} R_{12}^{uL}\lambda & 1 & R_{23}^{uL}\lambda^2  \\
\frac{1}{3} R_{23}^{uL}R_{12}^{uL}\lambda^3 & -R_{23}^{uL}\lambda^2 & 1 
\end{array}
\right),\label{eq:Lu}
\end{equation}
\begin{equation}
R_u \sim \left(
\begin{array}{ccc}
1 & -\frac{1}{3} R_{12}^{uL}\lambda & 0 \\
\frac{1}{3} R_{12}^{uL}\lambda & 1 & R_{23}^{uL}\lambda^2  \\
-\frac{1}{3} R_{23}^{uL}R_{12}^{uL}\lambda^3 & -R_{23}^{uL}\lambda^2 & 1 
\end{array}
\right),
\end{equation}
\begin{align}
L_d& =\left(
\begin{array}{cc}
1 \\
-(\frac{2}{3} R_{12}^{dL}+\frac{4}{27} I_{12}^{dL}e^{-i(2\rho - \delta)})\lambda \\
(-\frac{1}{3} R_{13}^{dL}+\frac{2}{3}R_{23}^{dL}R_{12}^{dL}+\frac{4}{27} R_{23}^{dL}I_{12}^{dL}
e^{-i(2\rho - \delta)})\lambda^3
\end{array}\right. \\
& \hspace{5.3cm} \left.
\begin{array}{cc}
(\frac{2}{3} R_{12}^{dL}+\frac{4}{27} I_{12}^{dL}e^{i(2\rho - \delta)})\lambda & 
\frac{1}{3} R_{13}^{dL}\lambda^3 \\
1 & R_{23}^{dL}\lambda^2 \\
-R_{23}^{dL}\lambda^2 & 1
\end{array}
\right), \nonumber
\end{align}
\begin{align}
R_d& =\left(
\begin{array}{cc}
1 & 
\frac{2}{3} I_{12}^{dR}e^{i(\rho-\delta)}\lambda^{0.5} \\
-\frac{2}{3} I_{12}^{dR}e^{-i(\rho-\delta)}\lambda^{0.5} & 1 \\
(-\frac{2}{3} R_{13}^{dR}+\frac{2}{3}I_{23}^{dR}I_{12}^{dR}-\frac{8}{27} I_{23}^{\prime dR}I_{12}^{dR}e^{-i(2\rho-\delta)})
\lambda & -I_{23}^{dR}e^{i(\rho-\delta)}\lambda^{0.5}
\end{array}\right. \nonumber \\
& \hspace{8.3cm} \left.
\begin{array}{c}
\frac{2}{3} R_{13}^{dR}\lambda \\
I_{23}^{dR}e^{-i(\rho-\delta)}\lambda^{0.5} \\
1
\end{array}
\right),
\end{align}

\begin{equation}
L_e \sim \left(
\begin{array}{ccc}
1 & I_{12}^{eL}e^{i(\rho-\delta)}\lambda^{0.5} & 0 \\
-I_{12}^{eL}e^{-i(\rho-\delta)}\lambda^{0.5} & 1 & I_{23}^{dR}e^{-i(\rho-\delta)}\lambda^{0.5}  \\
I_{23}^{dR}I_{12}^{eL}\lambda & -I_{23}^{dR}e^{i(\rho-\delta)}\lambda^{0.5} & 1 
\end{array}
\right),
\end{equation}
\begin{equation}
R_e \sim \left(
\begin{array}{ccc}
1 & R_{12}^{eR}\lambda & R_{13}^{eR}\lambda^3 \\
-R_{12}^{eR}\lambda & 1 & R_{23}^{dL}\lambda^2  \\
(-R_{13}^{eR}+R_{23}^{dL}R_{12}^{eR})\lambda^3 & -R_{23}^{dL}\lambda^2 & 1 
\end{array}
\right).\label{eq:Re}
\end{equation}
In this model the Majorana neutrino mass matrix has a lot of other real parameters and CP phases,
and therefore, we cannot constrain the diagonalizing matrix for neutrino. 
The diagonalizing matrix for neutrino is written as 
\begin{equation}
L_{\nu} \sim \left(
\begin{array}{ccc}
1 & \lambda^{0.5} & \lambda \\
\lambda^{0.5} & 1 & \lambda^{0.5}  \\
\lambda & \lambda^{0.5} & 1 
\end{array}
\right),
\end{equation}
where we have omitted the complex $O(1)$ coefficients. 
In this model we can obtain realistic CKM and MNS matrices as 
\begin{align}
U_{CKM} =L_u^{\dagger}L_d &\sim \left(
\begin{array}{c}
1    \\
(\frac{1}{3}R_{12}^{uL}-\frac{2}{3}R_{12}^{dL}-\frac{4}{27}I_{12}^{dL}e^{-i(2\rho-\delta)}) \lambda  \\
\{-\frac{2}{3}R_{23}^{uL}R_{12}^{dL}-\frac{1}{3}R_{13}^{dL}+\frac{2}{3}I_{23}^{dL}I_{12}^{dL}
-\frac{4}{27}(R_{23}^{uL}-R_{23}^{dL})I_{12}^{dL}e^{-i(2\rho-\delta)}\} \lambda^3 
\end{array} \right. \nonumber \\
& \hspace{-0.5cm} \left.
\begin{array}{cc} 
(-\frac{1}{3}R_{12}^{uL}+\frac{2}{3}R_{12}^{dL}+\frac{4}{27}I_{12}^{dL}e^{i(\rho-\delta)})\lambda & 
O(\lambda^4) \\ 
1 & 
(-R_{23}^{uL}+R_{23}^{dL})\lambda^2 \\
(R_{23}^{uL}-R_{23}^{dL})\lambda^2 & 
1
\end{array}\right),
\end{align}
\begin{equation}
\left| U_{MNS}  \right| = \left| L_{\nu}^{\dagger}L_e  \right| \sim \left(
\begin{array}{ccc}
1 & \lambda^{0.5} & \lambda \\
\lambda^{0.5} & 1 & \lambda^{0.5}  \\
\lambda & \lambda^{0.5} & 1 
\end{array}
\right).
\end{equation}
As discussed in Ref. \cite{SpCP,SpCPneutrino}, 
the leading contribution to the component $(U_{CKM})_{13}$
is cancelled and the subleading contribution $O(\lambda^4)$ dominates $(U_{CKM})_{13}$.

\section{Mass insertions}
\label{sec:mass insertions}
In this appendix, we just show all mass insertion parameters in this model.
\begin{equation}
(\delta_{12}^u)_{LL} = -(\delta_{12}^u)_{RR} \simeq \left\{ -\frac{1}{3}R_{12}^{uL} \lambda \Delta m^2_{10,2} 
-\frac{1}{3}(R_{23}^{uL})^2 R_{12}^{uL} \lambda^5 \Delta m^2_{10,3}\right\}/ m^2_{\tilde u}
\end{equation}
\begin{equation}
(\delta_{13}^u)_{LL} = -(\delta_{13}^u)_{RR} \simeq \left\{ -\frac{1}{3}R_{23}^{uL}R_{12}^{uL} \Delta m^2_{10,2} 
+\frac{1}{3}R_{23}^{uL} R_{12}^{uL} \Delta m^2_{10,3}\right\}\lambda^3/ m^2_{\tilde u}
\end{equation}
\begin{equation}
(\delta_{23}^u)_{LL} =(\delta_{23}^u)_{RR} \simeq R_{23}^{uL} \{ \Delta m^2_{10,2} 
- \Delta m^2_{10,3}\}\lambda^2/ m^2_{\tilde u}
\end{equation}
\begin{eqnarray}
&&(\delta_{12}^d)_{LL} \simeq \left\{-\left(\frac{2}{3}R_{12}^{dL}+\frac{4}{27}I_{12}^{dL}
e^{i(2\rho-\delta)}\right) \lambda \Delta m^2_{10,2}\right. \\
&&\hspace{2cm}\left.-R_{23}^{dL}\left(-\frac{1}{3}R_{13}^{dL}+\frac{2}{3}R_{23}^{dL}R_{12}^{dL}+
\frac{4}{27}R_{23}^{dL}I_{12}^{dL}e^{i(2\rho-\delta)}\right)\lambda^5 \Delta m^2_{10,3} \right\}
/ m^2_{\tilde d} \nonumber 
\end{eqnarray}
\begin{eqnarray}
&&(\delta_{13}^d)_{LL} \simeq \left\{ -R_{23}^{dL}\left(\frac{2}{3}R_{12}^{dL}+\frac{4}{27}I_{12}^{dL}
e^{i(2\rho-\delta)}\right) \Delta m^2_{10,2}\right. \\
&&\hspace{2cm}\left.+\left(-\frac{1}{3}R_{13}^{dL}+\frac{2}{3}R_{23}^{dL}R_{12}^{dL}+
\frac{4}{27}R_{23}^{dL}I_{12}^{dL}e^{i(2\rho-\delta)}\right) \Delta m^2_{10,3} \right\} \lambda^3
/ m^2_{\tilde d} \nonumber
\end{eqnarray}
\begin{equation}
(\delta_{23}^d)_{LL} \simeq  R_{23}^{dL} \{ \Delta m^2_{10,2} 
- \Delta m^2_{10,3} \} \lambda^2
/ m^2_{\tilde d} 
\end{equation}
\begin{equation}
(\delta_{12}^d)_{RR} \simeq \left\{ -\frac{2}{3}I_{12}^{dR} e^{i(\rho - \delta)} \lambda^{0.5} 
\Delta m^2_{\bar 5,2} 
-I_{23}^{dR}\left(-\frac{2}{3}R_{13}^{dR}+\frac{2}{3}I_{23}^{dR}I_{12}^{dR}\right)e^{i(\rho-\delta)}
\lambda^{1.5} \Delta m^2_{\bar 5,3} \right\} 
/ m^2_{\tilde d} 
\end{equation}
\begin{eqnarray}
&&(\delta_{13}^d)_{RR} \simeq \left\{ \left(-\frac{2}{3}I_{23}^{dR}I_{12}^{dR}
+\frac{8}{27}I_{12}^{dR}I_{23}^{\prime dR} e^{i(2\rho-\delta)}\right) 
\Delta m^2_{\bar 5,2}\right. \\
&&\hspace{2cm}\left.+\left(-\frac{2}{3}R_{13}^{dR}+\frac{2}{3}I_{23}^{dR}I_{12}^{dR}-
\frac{8}{27}I_{23}^{\prime dR}I_{12}^{dR}e^{i(2\rho-\delta)}\right) \Delta m^2_{\bar 5,3} \right\} \lambda
/ m^2_{\tilde d} \nonumber
\end{eqnarray}
\begin{equation}
(\delta_{23}^d)_{RR} \simeq I_{23}^{dR}e^{-i(\rho-\delta)}\{ 
\Delta m^2_{\bar 5,2} 
- \Delta m^2_{\bar 5,3} \} \lambda^{0.5}
/ m^2_{\tilde d} 
\end{equation}
\begin{equation}
(\delta_{12}^e)_{LL} \simeq -I_{12}^{eL}e^{i(\rho-\delta)}\{ 
\lambda^{0.5} \Delta m^2_{\bar 5,2} 
+ (I_{23}^{dR})^2 \lambda^{1.5} \Delta m^2_{\bar 5,3} \}
/ m^2_{\tilde e} 
\end{equation}
\begin{equation}
(\delta_{13}^e)_{LL} \simeq -I_{23}^{dR}I_{12}^{eL}\{
 \Delta m^2_{\bar 5,2} 
- \Delta m^2_{\bar 5,3} \} \lambda
/ m^2_{\tilde e} 
\end{equation}
\begin{equation}
(\delta_{23}^e)_{LL} \simeq I_{23}^{dR} e^{-i(\rho-\delta)} \{
 \Delta m^2_{\bar 5,2} 
- \Delta m^2_{\bar 5,3} \} \lambda^{0.5}
/ m^2_{\tilde e} 
\end{equation}
\begin{equation}
(\delta_{12}^e)_{RR} \simeq \{ -R_{12}^{eR}\lambda \Delta m^2_{10,2} 
- R_{23}^{dL}(-R_{13}^{eR}+R_{23}^{dL}R_{12}^{eR}) \lambda^5 \Delta m^2_{10,3} \}
/ m^2_{\tilde e} 
\end{equation}
\begin{equation}
(\delta_{13}^e)_{RR} \simeq \{ -R_{23}^{dL} R_{12}^{eR} \Delta m^2_{10,2} 
+(-R_{13}^{eR}+R_{23}^{dL}R_{12}^{eR}) \Delta m^2_{10,3} \} \lambda^3
/ m^2_{\tilde e} 
\end{equation}
\begin{equation}
(\delta_{23}^e)_{RR} \simeq R_{23}^{dL} \{
 \Delta m^2_{10,2} 
- \Delta m^2_{10,3} \} \lambda^2
/ m^2_{\tilde e} 
\end{equation}


\begin{thebibliography}{99}
\bibitem{GUT}
  H.~Georgi and S.~L.~Glashow,
  Phys.\ Rev.\ Lett.\  {\bf 32}, 438 (1974).

\bibitem{anarchy}
  L.~J.~Hall, H.~Murayama and N.~Weiner,
  Phys.\ Rev.\ Lett.\  {\bf 84}, 2572 (2000)
  [hep-ph/9911341];
  J.~Hisano, K.~Kurosawa and Y.~Nomura,
  Nucl.\ Phys.\ B {\bf 584}, 3 (2000)
  [hep-ph/0002286].

\bibitem{E6}
  H.~Fritzsch and P.~Minkowski,
  Annals Phys.\  {\bf 93}, 193 (1975);
  F.~Gursey, P.~Ramond and P.~Sikivie,
  Phys.\ Lett.\  B {\bf 60}, 177 (1976);
  Y.~Achiman and B.~Stech,
  Phys.\ Lett.\  B {\bf 77}, 389 (1978);
  R.~Barbieri and D.~V.~Nanopoulos,
  Phys.\ Lett.\  B {\bf 91}, 369 (1980).

\bibitem{E6neutrino}
  T.~Kugo and J.~Sato,
  Prog.\ Theor.\ Phys.\  {\bf 91}, 1217 (1994)
  [arXiv:hep-ph/9402357];
  N.~Irges, S.~Lavignac and P.~Ramond,
  Phys.\ Rev.\  D {\bf 58}, 035003 (1998)
  [arXiv:hep-ph/9802334];
%
  M.~Bando and T.~Kugo,
  Prog.\ Theor.\ Phys.\  {\bf 101}, 1313 (1999)
  [arXiv:hep-ph/9902204];
  M.~Bando, T.~Kugo and K.~Yoshioka,
  Prog.\ Theor.\ Phys.\  {\bf 104}, 211 (2000)
  [arXiv:hep-ph/0003220];
 
\bibitem{E6matter}
  M.~Bando and N.~Maekawa,
  Prog.\ Theor.\ Phys.\  {\bf 106}, 1255 (2001)
  [hep-ph/0109018].

\bibitem{family}
 M. Dine, A. Kagan, and R. Leigh, Phys. Rev. D {\bf 48},
                   4269 (1993) [hep-ph/9304299]; 
 A. Pomarol and D. Tommasini, Nucl. Phys. B {\bf 466}, 3 (1996) [hep-ph/9507462]; 
 R. Barbieri, G. Dvali, and L.J. Hall, Phys. Lett. B {\bf 377}, 76 (1996) [hep-ph/9512388];
 R. Barbieri and L.J. Hall, Nuovo Cim. A {\bf 110}, 1 (1997) [hep-ph/9605224];
 K.S. Babu and S.M. Barr, Phys. Lett. B {\bf 387}, 87 (1996) [hep-ph/9606384];
 R. Barbieri, L.J. Hall, S. Raby, and A. Romanino, Nucl. Phys. B {\bf 493}, 3 (1997)
 [hep-ph/9610449];
 Z. Berezhiani, Phys. Lett. B {\bf 417}, 287 (1998) [hep-ph/9609342];
 G. Eyal, Phys. Lett. B {\bf 441}, 191 (1998) [hep-ph/9807308];
 R. Barbieri, P. Creminelli, and A. Romanino, Nucl. Phys. B {\bf 559}, 17 (1999) 
 [hep-ph/9903460];
 S.F. King and G.G. Ross, Phys. Lett. B {\bf 520}, 243 (2001) [hep-ph/0108112].   

\bibitem{E6family}
  N.~Maekawa,
  Phys.\ Lett.\ B {\bf 561}, 273 (2003)
  [hep-ph/0212141];
  Prog.\ Theor.\ Phys.\  {\bf 112}, 639 (2004)
  [hep-ph/0402224];
  S.~-G.~Kim, N.~Maekawa, A.~Matsuzaki, K.~Sakurai and T.~Yoshikawa,
  Phys.\ Rev.\ D {\bf 75}, 115008 (2007)
  [hep-ph/0612370];
  S.~-G.~Kim, N.~Maekawa, A.~Matsuzaki, K.~Sakurai and T.~Yoshikawa,
  Prog.\ Theor.\ Phys.\  {\bf 121}, 49 (2009)
  [arXiv:0803.4250 [hep-ph]].
 
\bibitem{naturalSUSY}
A.~Cohen, D.~Kaplan and A.~Nelson, Phys. Lett. B {\bf 388}, 588 (1996) [hep-ph/9607394];
N.~Arkani-Hamed and H.~Murayama, Phys. Rev. D {\bf 56}, 6733 (1997) [hep-ph/9703259].

\bibitem{DT}
 For the review,
  L.~Randall and C.~Csaki,
  In *Palaiseau 1995, SUSY 95* 99-109
  [hep-ph/9508208].

\bibitem{anomalous}
   E.~Witten,  Phys. Lett. B {\bf 149} (1984),351;
                  M.~Dine, N.~Seiberg and E.~Witten,
                  Nucl. Phys. B {\bf 289} (1987), 589;
                  J.J.~Atick, L.J.~Dixon and A.~Sen,
                  Nucl. Phys. B {\bf 292} (1987),109;
                  M.~Dine, I.~Ichinose and N.~Seiberg,
                  Nucl. Phys. B {\bf 293} (1987),253.

\bibitem{naturalGUT}
  N.~Maekawa,
  Prog.\ Theor.\ Phys.\  {\bf 106}, 401 (2001)
  [hep-ph/0104200];
  Prog.\ Theor.\ Phys.\  {\bf 107}, 597 (2002)
  [hep-ph/0111205];
  N.~Maekawa and T.~Yamashita,
  Phys.\ Rev.\ Lett.\  {\bf 90}, 121801 (2003)
  [hep-ph/0209217];
  Prog.\ Theor.\ Phys.\  {\bf 107}, 1201 (2002)
  [hep-ph/0202050];
  Prog.\ Theor.\ Phys.\  {\bf 110}, 93 (2003)
  [hep-ph/0303207].

\bibitem{mu} 
  N.~Maekawa,
  Phys.\ Lett.\ B {\bf 521}, 42 (2001)
  [hep-ph/0107313].

\bibitem{non-GUT mu}
  R.~Hempfling, Phys. Lett.\ B {\bf329}, 222 (1994) [hep-ph/9404257];

\bibitem{CEDM}
  J.~Hisano and Y.~Shimizu,
  Phys.\ Rev.\  D {\bf 70}, 093001 (2004)
  [arXiv:hep-ph/0406091];
  K.~V.~P.~Latha, D.~Angom, B.~P.~Das and D.~Mukherjee,
  arXiv:0902.4790 [physics.atom-ph];
  J.~R.~Ellis, J.~S.~Lee and A.~Pilaftsis,
  JHEP {\bf 0810}, 049 (2008)
  [arXiv:0808.1819 [hep-ph]].

\bibitem{SpCP}
  M.~Ishiduki, S.~-G.~Kim, N.~Maekawa and K.~Sakurai,
  Prog.\ Theor.\ Phys.\  {\bf 122}, 659 (2009)
  [arXiv:0901.3400 [hep-ph]];
  Phys.\ Rev.\ D {\bf 80}, 115011 (2009)
  [Erratum-ibid.\ D {\bf 81}, 039901 (2010)]
  [arXiv:0910.1336 [hep-ph]];
@  H.~Kawase and N.~Maekawa,
  Prog.\ Theor.\ Phys.\  {\bf 123}, 941 (2010)
  [arXiv:1005.1049 [hep-ph]].

\bibitem{SpCPneutrino}
  N.~Maekawa and K.~Takayama,
  Phys.\ Rev.\ D {\bf 85}, 095015 (2012)
  [arXiv:1202.5816 [hep-ph]].

\bibitem{Proton}
N.~Maekawa and Y.~Muramatsu,
  Phys.\ Rev.\ D {\bf 88}, 095008 (2013)
  [arXiv:1307.7529].

\bibitem{ProtonSU2}
  N.~Maekawa and Y.~Muramatsu,
  arXiv:1401.2633 [hep-ph].
 
\bibitem{kawamura}
  J.~S.~Hagelin and S.~Kelley,
  Nucl.\ Phys.\ B {\bf 342}, 95 (1990);
  Y.~Kawamura, H.~Murayama and M.~Yamaguchi,
  Phys.\ Lett.\ B {\bf 324}, 52 (1994)
  [hep-ph/9402254];
  Phys.\ Rev.\ D {\bf 51}, 1337 (1995)
  [hep-ph/9406245];
  Y.~Kawamura and M.~Tanaka,
  Prog.\ Theor.\ Phys.\  {\bf 91}, 949 (1994).

\bibitem{Higgs}  
  G.~Aad {\it et al.}  [ATLAS Collaboration],
  Phys.\ Lett.\ B {\bf 716}, 1 (2012)
  [arXiv:1207.7214 [hep-ex]];
  S.~Chatrchyan {\it et al.}  [CMS Collaboration],
  Phys.\ Lett.\ B {\bf 716}, 30 (2012)
  [arXiv:1207.7235 [hep-ex]].

  
\bibitem{yoshioka}
  K.~Inoue, K.~Kojima and K.~Yoshioka,
  JHEP {\bf 0707}, 027 (2007)
  [hep-ph/0703253].
 
\bibitem{FCNC}
  F.~Gabbiani, E.~Gabrielli, A.~Masiero and L.~Silvestrini,
  Nucl.\ Phys.\ B {\bf 477}, 321 (1996)
  [hep-ph/9604387];
  M.~Ciuchini, V.~Lubicz, L.~Conti, A.~Vladikas, A.~Donini, E.~Franco, G.~Martinelli and I.~Scimemi {\it et al.},
  JHEP {\bf 9810}, 008 (1998)
  [hep-ph/9808328].

\bibitem{takayama}
  N.~Maekawa and K.~Takayama,
  arXiv:1403.7629 [hep-ph].


\end{thebibliography}
\end{document}